\documentclass[twocolumn,pre,floats,aps,amsmath,amssymb,nofootinbib]{revtex4-1}
\usepackage{graphicx}
\usepackage{bm}
\usepackage{physics}

\begin{document}

\title{No Paradox in Wave-Particle Duality}
\author{Andrew Knight}
\affiliation{aknight@alum.mit.edu}
\date{\today}

\begin{abstract}
The assertion that an experiment by Afshar \textit{et al.}\ demonstrates violation of Bohr’s Principle of Complementarity is based on the faulty assumption that which-way information in a double-slit interference experiment can be retroactively determined from a future measurement.
\end{abstract}

\maketitle

Afshar \textit{et al.}\ \cite{Afshar} claim to demonstrate a violation of Bohr's Principle of Complementarity (``BPC") in a double-slit experiment in which, they claim, both which-way information and interference visibility are available for the same photons.  They assert that the experiment simultaneously yields distinguishability D of which-way information and visibility V of interference that violate the complementarity inequality $D^2 + V^2 \leq 1$. 

In the experiment, a laser beam irradiates adjacent pinholes A and B and the emergent light is focused via a converging lens located in the far field onto detectors 1 and 2.  The experiment is set up so that when pinhole B is closed, detector 1 detects essentially all photons, while when pinhole A is closed, detector 2 detects essentially all photons.  Further, thin wires are placed in the Fourier plane, directly in front of the lens, where destructive interference would be expected.  As a result, the authors of \cite{Afshar} are able to correctly infer the existence of interference when both pinholes are open because the wires have an insignificant effect on the total radiation detected by the detectors.\footnote{With both pinholes open, the wires reduce the detected intensity by about 2\%, consistent with the wires located at interference minima; with one pinhole open, the wires reduce the detected intensity by about 15\%, consistent with the absence of interference.}  Their experimental setup is clever and their contribution to physics -- demonstrating the presence of quantum interference by the lack of a significant reduction of the passing light due to the thin wires -- is important.  Nevertheless, they go too far: their conclusion that BPC is violated depends crucially on their claim to have ``had full which-way information when... two pinholes were open...".\footnote{To support this notion, they cite \cite{Wheeler} and \cite{Bartell}.}  This claim is incorrect.

In Fig.\ 1, a beam of light L is shown at time $t_0$ that is spatially coherent over width W.\footnote{We can, as in \cite{Afshar}, stipulate that the photon flux is adequately low that we only consider one photon at a time.  Because each photon is spatially coherent over width W, it is imperative not to regard it as being located somewhere specific within W, but rather that no information exists to distinguish its location anywhere within W.  Further, the photon need not be produced by a laser to be spatially coherent \cite{Pearson}.  However, the use in \cite{Afshar} of a laser beam (whose width necessarily exceeds the width spanned by the pinholes) to irradiate the pinholes guarantees that it is.}  Let us initially ignore any event at time $t_1$.  At time $t_2$, a photon is detected -- that is, it is localized to within some dimension that is smaller than W.  A simple question: \textbf{does the photon's localization at $t_2$ provide information about its specific location within width W at time $t_0$?}\footnote{Alternatively: does decoherence at $t_2$ retroactively change any facts about the beam at $t_0$?}

\begin{figure}[ht]
\includegraphics[width=3.2 in]{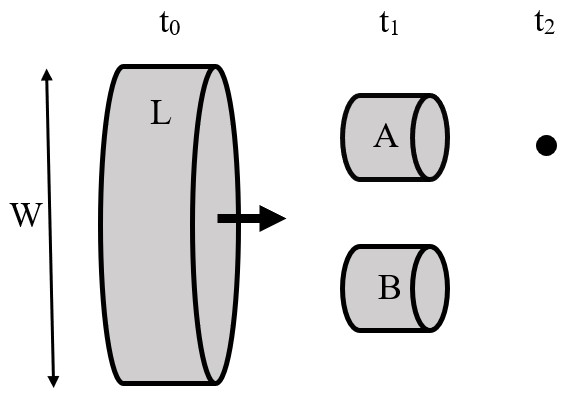}
\caption{Beam of light L, spatially coherent over width W at time $t_0$, passes through pinholes A and B at time $t_1$ before detection of a photon at time $t_2$.}
\end{figure}

No.  The correct answer to this question is so fundamental to the foundations of physics and quantum mechanics that any other response inevitably leads to paradox and confusion.\footnote{One might argue that in the de Broglie-Bohm interpretation of quantum mechanics, the correct answer to this question might not be so clear.  However, such an objection is irrelevant to this paper for several reasons.  First, as an interpretation of quantum mechanics, Bohmian mechanics is empirically indistinguishable from other interpretations, thus it provides no scientific means to counter the assertion that the photon's localization at $t_2$ provides no information about its location within W at $t_0$.  Second, because each particle in Bohmian mechanics always has a well-defined position that deterministically changes according to a pilot wave (which depends on the quantum wave state), its particle-like and wave-like behaviors are not constrained by complementarity.  In other words, Bohmian mechanics inevitably denies BPC anyway, rendering moot the question of whether there is anything special about the experimental setup of \cite{Afshar} that demonstrates violation of BPC.}  The quantum wave state of an object -- written in the position basis -- provides the ability to predict the likelihood of finding the object in a particular position in a future ``measurement."  Such a measurement broadly includes any interaction that correlates the object to the environment, such as entanglement with a photon that reflects off the object.  Until such an interaction occurs, the wave state evolves linearly with time.  Thus, if the wave state of the object is known just prior to a measurement at time $t_0$, a probabilistic prediction can be made about what value the measurement will yield.  However, if no measurement is made until a later time $t_2$, the measured value does not retroactively determine what value \textit{would have} been measured at time $t_0$.  In other words, there is no fact about where the object would have been detected if it was not, in fact, detected.

In the same vein, in Fig.\ 1, there is no fact about where (within width W) the photon was located at time $t_0$ if it was, in fact, spatially coherent over that width.  Indeed, the preparation of the beam to be spatially coherent over width W at $t_0$ guarantees that there is no fact about the photon's specific position within width W -- i.e., quantum mechanically, the state of the photon can be represented as a superposition over all positions within W.\footnote{The Fourier transform of this state will yield a superposition over all possible momenta.  The narrower W, the wider the spread of possible momenta \textit{a la} quantum uncertainty, which is why narrow laser beams disperse at a larger angle than wider beams.}  It may be tempting to look at the photon's detected position at $t_2$ and then infer, based on conservation of momentum or other considerations, where it ``must have been" at $t_0$, or to trace out a path that it ``must have taken," but such a retroactive localization conflicts with the photon's known spatial coherence at $t_0$.  Logically, such backward-in-time inferences are incompatible with the existence of quantum superpositions, without which we would not observe interference.

However, Ref.\ \cite{Afshar} inherently assumes the answer to this fundamental question is yes.\footnote{``[T]he complementary measurements refer back to what `takes place' at the pinholes when a photon passes that plane."}  To show this, let us introduce in Fig.\ 1 a ``double-slit" at time $t_1$ comprising a material that allows the photon to pass only through pinholes A and B.  The analysis is no different.  Ref.\ \cite{Afshar} notes that pinhole A is correlated with detector 1 when pinhole B is closed and that pinhole B is correlated with detector 2 when pinhole A is closed.  It is understandably tempting to then infer, as \cite{Afshar} does, that these correlations remain when both pinholes are open.  But this is exactly the kind of inference that is not allowed by quantum mechanics because they are different experiments with different measurements.  With both pinholes open, measurement by the detectors at $t_2$ tells us nothing about which of the two pinholes the photon traversed at $t_1$, for the same reason that, with or without the ``double slit," localization of the photon at $t_2$ tells us nothing about where the photon ``was" at $t_0$.  As tempting as it may be to apply classical reasoning in Fig.\ 1 to conclude that the photon detected at $t_2$ must have passed through pinhole A at $t_1$, such a conclusion is unfounded.  If there is no event/measurement/correlation at $t_1$ that can distinguish pinhole A from B, then no future event/measurement/correlation will retroactively distinguish them.

With both pinholes open, the pinholes have the effect of localizing the original photon, which was spatially coherent over a width wider than that spanned by the two pinholes, to within the regions defined by the two pinholes.  If the photon reaches the pinholes at $t_1$, then, neglecting complex coefficients, its wave function $\Psi(t_1) = \Psi_A(t_1) + \Psi_B(t_1)$, where $\Psi_A(t_1)$  is the wave function that would have described the photon at $t_1$ if pinhole B had been closed and $\Psi_B(t_1)$  is the wave function that would have described the photon at $t_1$ if pinhole A had been closed.  Knowledge of the time evolutions of $\Psi_A(t)$ and $\Psi_B(t)$ provides the powerful ability to probabilistically predict where the photon might be detected in the future.  However,  Ref.\ \cite{Afshar} asserts that a measurement at $t_2$ allows us to retroactively conclude that $\Psi(t_1) = \Psi_A(t_1)$ \textit{or} $\Psi_B(t_1)$.  But this clearly contradicts the experimental setup in which a spatially coherent beam is incident on the two pinholes.  Even if Afshar \textit{et al.}\ didn't immediately recognize the problem with retrocausality, the fact that they actually observed interference in the far field of the two pinholes conflicts with localization of the photon in either pinhole A or B at $t_1$.  With both pinholes open, there simply is no fact about the passage of the photon through either pinhole A or B, a necessary result of the experimental setup and readily confirmed by the presence of interference in the far field.  Because the experiment was set up so that each photon was spatially coherent (and thus indistinguishable) over pinholes A and B, then no ``which-way" information can ever exist and no future measurement can retroactively distinguish them.

In other words, Afshar \textit{et al.} claim in one breath to have set up the experiment so that pinholes A and B are inherently indistinguishable by certain photons\footnote{Specifically, photons that are produced to be spatially coherent over the width spanned by pinholes that are thus incapable of distinguishing them.}, and in another breath to have distinguished pinholes A and B with those same photons.  We do not need hundreds of pages of complicated logical analysis and several dozen mathematical equations to see that these claims are self-contradictory -- yet, that is exactly what we get in the citing literature, of which no fewer than a dozen publications debate the correctness of \cite{Afshar}.  While the majority of these refute \cite{Afshar} (e.g., \cite{Kastner,St,Jacques,Georgiev,Drezet,Qureshi}, but see \cite{Flores1,Flores2}), a variety of conflicting reasons is given \cite{Kaloyerou}, leading some publications to \textit{incorrectly} refute \cite{Afshar}.  Ultimately, \cite{Afshar} has been cited by over a hundred publications, very few of which refute it, and many of which cite it to support incorrect conclusions.

Ref.\ \cite{Afshar} is problematic in the foundations of physics.  What makes it a problem is not that it is incorrect; science could never progress if scientists were not allowed to make mistakes.  Further, its experimental setup is unique and elucidating.  The real problem with \cite{Afshar} is that it continues to be cited and taken seriously for its proposition that its experimental setup violates BPC and allows a superposition to carry which-way information, thus continuing to muddle the ether of already confusing and conflicting characterizations of quantum mechanics.

\end{document}